\documentclass[aps,prappl,reprint,longbibliography,superscriptaddress]{revtex4-1}

\usepackage{graphicx}
\usepackage{dcolumn}
\usepackage{bm}
\usepackage{epsfig}
\usepackage{amsmath}
\usepackage{mathtools}
\usepackage{subfigure}
\usepackage{epstopdf}

\usepackage{url}
\usepackage[colorlinks]{hyperref}

\hypersetup{%
        plainpages=true,
        breaklinks=true,
        hypertexnames=false,
        pageanchor=true,
        colorlinks=true,
        linkcolor={blue},
        citecolor={magenta},
        urlcolor={blue},
        anchorcolor={black}
      }

\usepackage{mathrsfs,bm,amsthm,amsfonts,xspace,float,bookmark}
\usepackage{textcomp}
\usepackage{tabu}
\usepackage{bbold}
\usepackage{xspace}
\usepackage{float}
\usepackage[inline]{enumitem}

\newcommand{\ket}[1]{|#1\rangle}
\newcommand{\bra}[1]{\langle#1|}

\begin{document}

\author{Yong Lu}
\email[e-mail:]{yongl@chalmers.se}
\affiliation{Department of Microtechnology and Nanoscience (MC2), Chalmers University of Technology, SE-412 96 G\"{o}teborg, Sweden}

\author{Andreas Bengtsson}
\affiliation{Department of Microtechnology and Nanoscience (MC2), Chalmers University of Technology, SE-412 96 G\"{o}teborg, Sweden}

\author{Jonathan J. Burnett}
\affiliation{Department of Microtechnology and Nanoscience (MC2), Chalmers University of Technology, SE-412 96 G\"{o}teborg, Sweden}
\affiliation{National Physical Laboratory, Hampton Road, Teddington, Middlesex, TW11 0LW, United Kingdom}

\author{Emely Wiegand}
\affiliation{Department of Microtechnology and Nanoscience (MC2), Chalmers University of Technology, SE-412 96 G\"{o}teborg, Sweden}

\author{Baladitya Suri}
\affiliation{Department of Microtechnology and Nanoscience (MC2), Chalmers University of Technology, SE-412 96 G\"{o}teborg, Sweden}
\affiliation{Indian Institute of Science, Department of Instrumentation and Applied Physics, Bangalore 560012, India}

\author{Philip Krantz}
\affiliation{Department of Microtechnology and Nanoscience (MC2), Chalmers University of Technology, SE-412 96 G\"{o}teborg, Sweden}

\author{Anita Fadavi Roudsari}
\affiliation{Department of Microtechnology and Nanoscience (MC2), Chalmers University of Technology, SE-412 96 G\"{o}teborg, Sweden}

\author{Anton Frisk Kockum}
\affiliation{Department of Microtechnology and Nanoscience (MC2), Chalmers University of Technology, SE-412 96 G\"{o}teborg, Sweden}

\author{Simone Gasparinetti}
\affiliation{Department of Microtechnology and Nanoscience (MC2), Chalmers University of Technology, SE-412 96 G\"{o}teborg, Sweden}

\author{G\"{o}ran Johansson}
\affiliation{Department of Microtechnology and Nanoscience (MC2), Chalmers University of Technology, SE-412 96 G\"{o}teborg, Sweden}

\author{Per Delsing}
\email[e-mail:]{per.delsing@chalmers.se}
\affiliation{Department of Microtechnology and Nanoscience (MC2), Chalmers University of Technology, SE-412 96 G\"{o}teborg, Sweden}

\title{Determination of decay rates in waveguide quantum electrodynamics}
\title{Decoherence benchmarking of a superconducting qubit in waveguide quantum electrodynamics}
\title{Decoherence benchmarking of superconducting qubits from microwave scattering}
\title{Characterizing decoherence rates of a superconducting qubit\\by direct microwave scattering}
\begin{abstract}
We experimentally investigate a superconducting qubit coupled to the end of an open transmission line, in a regime where the qubit decay rates to the transmission line and to its own environment are comparable. We perform measurements of coherent and incoherent scattering, on- and off-resonant fluorescence, and time-resolved dynamics to determine the decay and decoherence rates of the qubit. In particular, these measurements let us discriminate between non-radiative decay and pure dephasing. We combine and contrast results across all methods and find consistent values for the extracted rates. The results show that the pure dephasing rate is one order of magnitude smaller than the non-radiative decay rate for our qubit. Our results indicate a pathway to benchmark decoherence rates of superconducting qubits in a resonator-free setting.

\end{abstract}
\maketitle

\section{INTRODUCTION}
\label{sec1}

Superconducting circuits are promising building blocks for implementing quantum computers~\cite{steffen2011quantum,arute2019quantum,barends2014superconducting}. In those devices, the key elements are superconducting artificial atoms made by Josephson junctions which induce a strong and engineerable nonlinearity. Such artificial atoms are also used in the field of superconducting waveguide quantum electrodynamics (waveguide QED)~\cite{gu2017microwave,roy2017colloquium}, where they interact with a continuum of light modes in a 1D waveguide. In the past decade, many quantum effects from atomic physics and quantum optics have been demonstrated in waveguide QED, e.g., the Mollow triplet~\cite{astafiev2010resonance}, giant cross-Kerr effect \cite{hoi2013giant} and cooperative effects~\cite{roy2017colloquium,van2013photon,mirhosseini2019cavity}. Other recent experiments have shown phenomena which are currently beyond the reach of atomic physics, such as ultra-strong~\cite{kockum2019ultrastrong,forn2017ultrastrong} and superstrong coupling~\cite{kuzmin2019superstrong} between light and matter. Waveguide QED is also an enabling quantum technology. One of the key applications is to generate~\cite{kuhn2002deterministic, motes2015linear, zhou2019tunable,peng2016tuneable,forn2017demand,pechal2016superconducting,gasparinetti2017correlations} and detect~\cite{gu2017microwave,fan2013breakdown,sathyamoorthy2014quantum,inomata2016single,kono2018quantum,royer2018itinerant,sathyamoorthy2016detecting,besse2018single} single photons. It has been proposed to use waveguide QED to create bound states~\cite{zheng2010waveguide,sanchez2017dynamical,calajo2019exciting} and implement quantum computers~\cite{paulisch2016universal,zheng2013waveguide,knill2001scheme}.

The performance of quantum computers and waveguide-QED devices is often limited by the coherence of the Josephson circuits. For example, the efficiency of producing and detecting single photons, the lifetime of bound states, and the fidelity of logical gates can all be improved by enhancing the coherence. In a waveguide-QED setup, decoherence can be due to decay into the waveguide, pure dephasing, and non-radiative decay rate into other modes. However, the rates for pure dephasing and non-radiative decay are typically not explored separately. An understanding of which one is dominant will give an insight into the decoherence mechanisms, and thus how device performance can be improved.

In this work, we probe a superconducting transmon qubit coupled directly to the end of an open transmission line. In previous realizations~\cite{Wen2019,wen2018reflective,astafiev2010resonance, hoi2015probing,hoi2013giant,van2013photon}, the coupling rates were much larger than intrinsic decoherence mechanisms of the qubit, so the effects of non-radiative decay and pure dephasing were small and could not be well characterized. Here, we investigate a qubit whose radiative decay rate into the transmission line is larger than, yet comparable to, other decoherence mechanisms. This allows us to explore the pure dephasing rate $\Gamma_{\phi}$, the radiative decay rate $\Gamma_{\rm{r}}$ from the capacitive coupling to the waveguide, and the non-radiative decay rate $\Gamma_{\rm{n}}$. The total relaxation and decoherence rates are given by $\Gamma_1 = \Gamma_{\rm{r}} + \Gamma_{\rm{n}}$ and $\Gamma_2 = \Gamma_1/2 + \Gamma_{\phi}$, respectively. We demonstrate different methods to extract the different rates and find consistent results. In contrast to the results in circuit
QED~\cite{muller2015interacting,klimov2018fluctuations,burnett2019decoherence,schlor2019correlating,dunsworth2017characterization}, our methods enable the evaluation of the decoherence of qubits over a broad range of frequencies, and provide a pathway to investigate Josephson junctions or superconducting quantum interference devices (SQUIDs) without any resonator. In addition, we also consider it important to study $\Gamma_n$ and $\Gamma_\phi$ separately. For instance, this could help to improve the Purcell enhancement factor, $\frac{\Gamma_r}{\Gamma_n+2\Gamma_{\phi}}$, in devices such as that presented in Ref.~\cite{mirhosseini2019cavity}. Moreover, the spontaneous-emission factor $\beta$, which is customarily quoted in other waveguide-QED platforms \cite{rao2007single,chu1993spontaneous,lecamp2007very,baba1991spontaneous} is also related to $\Gamma_n$, namely, $\beta=\frac{\Gamma_r}{\Gamma_r+\Gamma_n}$ in our case.

The paper is structured as follows: in Sec.~\uppercase\expandafter{\romannumeral2}, we characterize the coherent scattering of the device and obtain the radiative decay rate and the decoherence rate of the qubit as a reference for later measurements. In Sec.~\uppercase\expandafter{\romannumeral3}, we exploit the fluorescence of the qubit under coherent excitation to find the non-radiative decay rate and the pure dephasing rate. The resonance fluorescence spectrum at strong driving develops into the Mollow triplet \cite{mollow1969power}, which  has been widely used to probe quantum properties in systems based on superconducting qubits such as coherence \cite{astafiev2010resonance,van2013photon} and vacuum squeezing \cite{toyli2016resonance}. The resonance spectrum is symmetric around the central peak.  However, if pure dephasing exists, the off-resonant spectrum becomes asymmetric, something which has been studied experimentally in quantum dots~\cite{ulrich2011dephasing,roy2011phonon}. We take advantage of this fact to extract the pure dephasing rate. In Sec.\,\uppercase\expandafter{\romannumeral4}, we measure the non-radiative decay rate under a continuous coherent drive, where coherently and incoherently scattered photons provide information about the different decay channels. In Sec.~\uppercase\expandafter{\romannumeral5}, we apply a pulse to the qubit to both obtain the decay rates and find the stability of the qubit frequency and coherence as a function of time. In contrast to other methods, we use the phase information of emitted photons from the qubit to investigate the qubit-frequency stability in superconducting waveguide QED. Finally, in Sec.\,\uppercase\expandafter{\romannumeral6}, we summarize the measured results and compare the advantages and disadvantages of the different methods.

\section{Device characterization}
\label{sec1}
\begin{figure}[tbph]
\includegraphics[width=\linewidth]{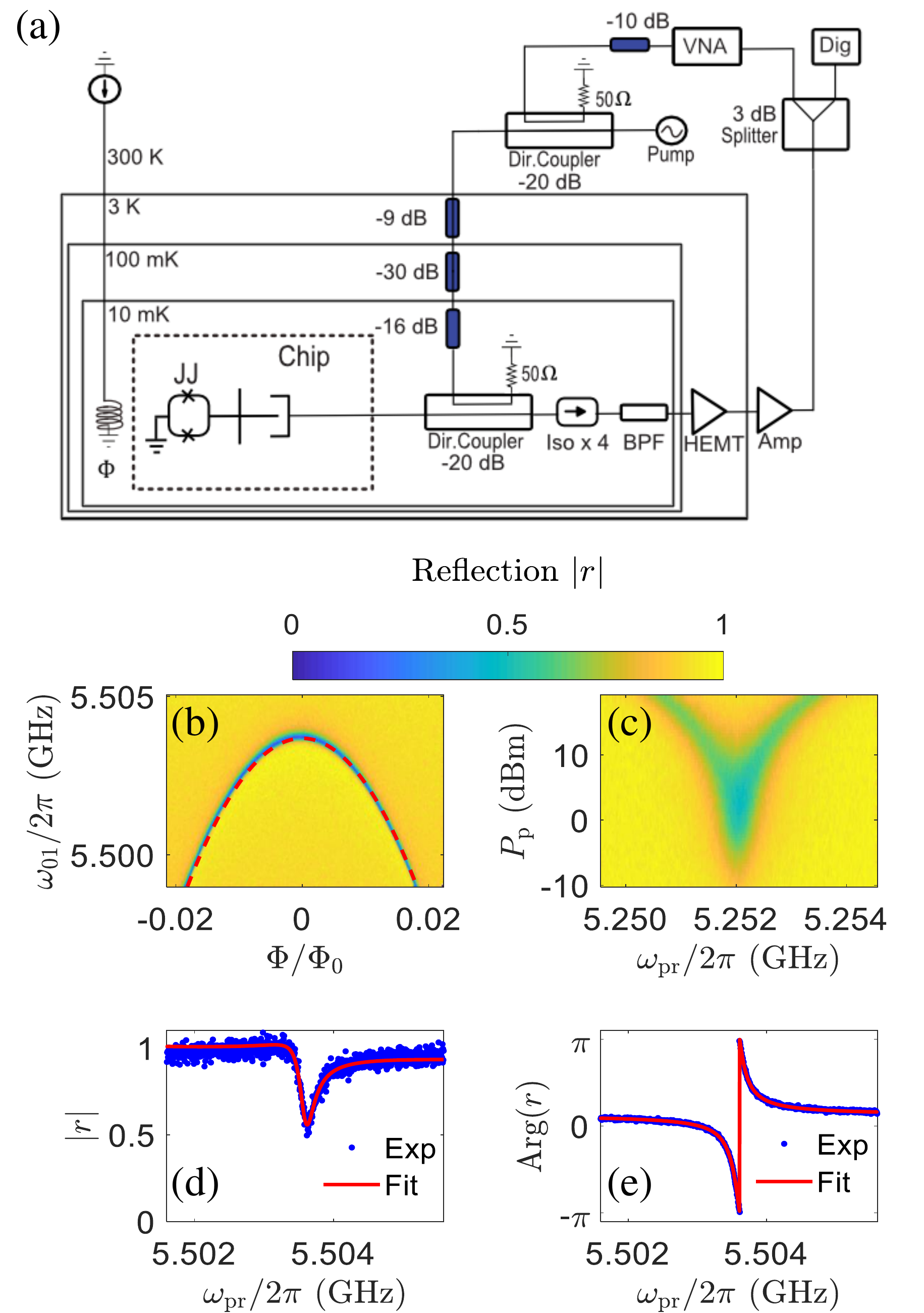}
\caption{Measurement setup and spectroscopy of a transmon qubit.
(a) A simplified schematic of the setup and experimental device. JJ, Iso, BPF, HEMT, Amp and Dig denote Josephson junctions, isolators, a bandpass filter, a high electron mobility transistor amplifier, room-temperature amplifiers, and a digitizer, respectively. In the dashed box is our chip, where a qubit is formed by a cross-shaped island connected to the ground plane via two Josephson junctions. The qubit is located at the end of an open transmission line with a coplanar geometry. A weak probe signal generated by a VNA is combined with a pump, using a directional coupler, fed through attenuators to the qubit in a cryostat cooled to 10\,\rm{mK}. The reflected signal is then measured by the VNA.
(b) Single-tone spectroscopy. The magnitude of the reflection coefficient $r$ is measured as a function of the external flux $\Phi$ and probe frequency $\omega_{\rm{pr}}$. The red dashed curve is a fit for the qubit frequency $\omega_{01}$.
(c) Two-tone spectroscopy. A strong pump is applied to the $|0\rangle-|1\rangle$ transition in order to saturate the population of the first excited state of the qubit. When the applied weak probe is on resonance with the
$|1\rangle-|2\rangle$ transition, the signal is either scattered incoherently or lost into the environment, leading to a reflection-coefficient magnitude less than unity. At higher 
pump power, the Autler$-$Townes splitting is observed \cite{autler1955stark}. From this, we obtain the qubit anharmonicity $\alpha=-252\,\rm{MHz}$.
(d) and (e) show the magnitude and phase response of the qubit at $\Phi=0$ under weak probing. Red lines are the corresponding fits using the circle fit technique from Ref.~\cite{probst2015efficient}.
}.
\label{setup}
\end{figure}
The device used in our experiment [see Fig.~\ref{setup}(a)] is a magnetic-flux-tunable Xmon-type transmon qubit~\cite{koch2007charge}, capacitively coupled to the open end of a one-dimensional transmission line with characteristic impedance $Z_0\simeq 50\,\Omega$. The circuit is equivalent to an atom in front of a mirror in 1D space. The device is fabricated from aluminum on a silicon substrate using the same fabrication recipe as in Ref.~\cite{burnett2019decoherence}. We denote  $|0\rangle$, $|1\rangle$ and $|2\rangle$ as ground state, first and second excited states of the qubit, respectively. The $|0\rangle-|1\rangle$ transition energy is $\hbar\omega_{01}\approx\sqrt{8E_J(\Phi)E_C}-E_C$, where $E_C=e^2/(2C_{\scriptscriptstyle\sum})$ is the charging energy, $\textit{e}$ is the elementary charge, $C_{{\scriptscriptstyle\sum}}$ is the total capacitance of the qubit, and $E_J(\Phi)$ is the Josephson energy. The Josephson energy can be tuned from its maximum value $E_{J,max}$ by an external magnetic flux $\Phi$ using a coil: $E_J(\Phi)=E_{J,max} |\cos(\pi\Phi/\Phi_0)|$, where $\Phi_0=h/(2e)$ is the magnetic flux quantum.

Figure~\ref{setup}(a) illustrates the simplified experimental setup for measuring the reflection coefficient of a probe signal from a vector network analyzer (VNA) after interacting with the qubit. The probe signal at frequency $\omega_{\rm{pr}}$ and a pump at frequency $\omega_{\rm{p}}$ are combined and attenuated before being fed into the transmission line. Then, the VNA receives the reflected signal to determine the complex reflection coefficient.

Figure~\ref{setup}(b) shows the magnitude of the reflection coefficient, $|r|$, for a weak probe (with an intensity $\Omega_{\rm{pr}}<\Gamma_2$) as a function of the external flux $\Phi$. We use two-tone spectroscopy to determine the anharmonicity of the qubit, $\alpha=(\omega_{12}-\omega_{01})/\hbar$ , where $\omega_{12}$ is the frequency of the $|1\rangle\leftrightarrow|2\rangle$ transition. Specifically, we apply a strong pump (with an intensity $\Omega_{\rm{p}}\gg \Gamma_2$) at $\omega_{01}$ to saturate the $|0\rangle-|1\rangle$ transition  and measure the reflection coefficient as a function of probe frequency. The result is shown in Fig.~\ref{setup}(c): a dip appears in the reflection at
$\omega_{\rm{pr}}=\omega_{12}$ due to the photon scattering from the $|1\rangle-|2\rangle$ transition. From Fig.~\ref{setup}(b) and (c),  we find $E_C \approx\alpha=252\,\rm{MHz}$, and then, $E_{J,max}=16.56\,\rm{GHz}$ by fitting the data in Fig.~\ref{setup}(b) to the equation between the qubit frequency and the external flux mentioned previously. In order to obtain the radiative decay and decoherence rates, we perform single-tone spectroscopy with a weak probe ($\Omega_{\rm{pr}}\ll\Gamma_2$). Figure~\ref{setup}(d) and (e) are the corresponding magnitude and phase response of $r$, where we obtain $\Gamma_{\rm{r}}/2\pi=227\,\rm{kHz}$ and $\Gamma_2/2\pi=141\,\rm{kHz}$ by using the circle fit technique from Ref.~\cite{probst2015efficient}.
\section{Atomic fluorescence}
Even though a measurement of the reflection coefficient can give the decoherence and radiative decay rates, it cannot distinguish between pure dephasing and non-radiative decay. In order to distinguish them, we study the atomic fluorescence for different pump intensities and frequencies. For this measurement, the VNA is turned off, the pump is used to drive the system, another 50\,\rm{dB} of attenuation is added between the directional coupler and the pump, and the output signal is sampled by a digitizer [compare Fig.~\ref{setup}(a)].  When the qubit is pumped, its state evolves at a Rabi frequency $\Omega$. With a Rabi frequency much larger than the natural linewidth of the qubit ($\Omega\gg\Gamma_2$), the energy levels of the qubit become  dressed, leading to three distinct spectral components known as the Mollow triplet~\cite{mollow1969power}. In particular, the spectrum contains the elastic 'Rayleigh' line in the middle in which the scattered wave has the same frequency as the incident wave, with two inelastic sidebands positioned symmetrically on both sides of the center peak.
\subsection{On-resonant Mollow triplet}
\label{sec2}
\begin{figure}[tbph]
\includegraphics[width=\linewidth]{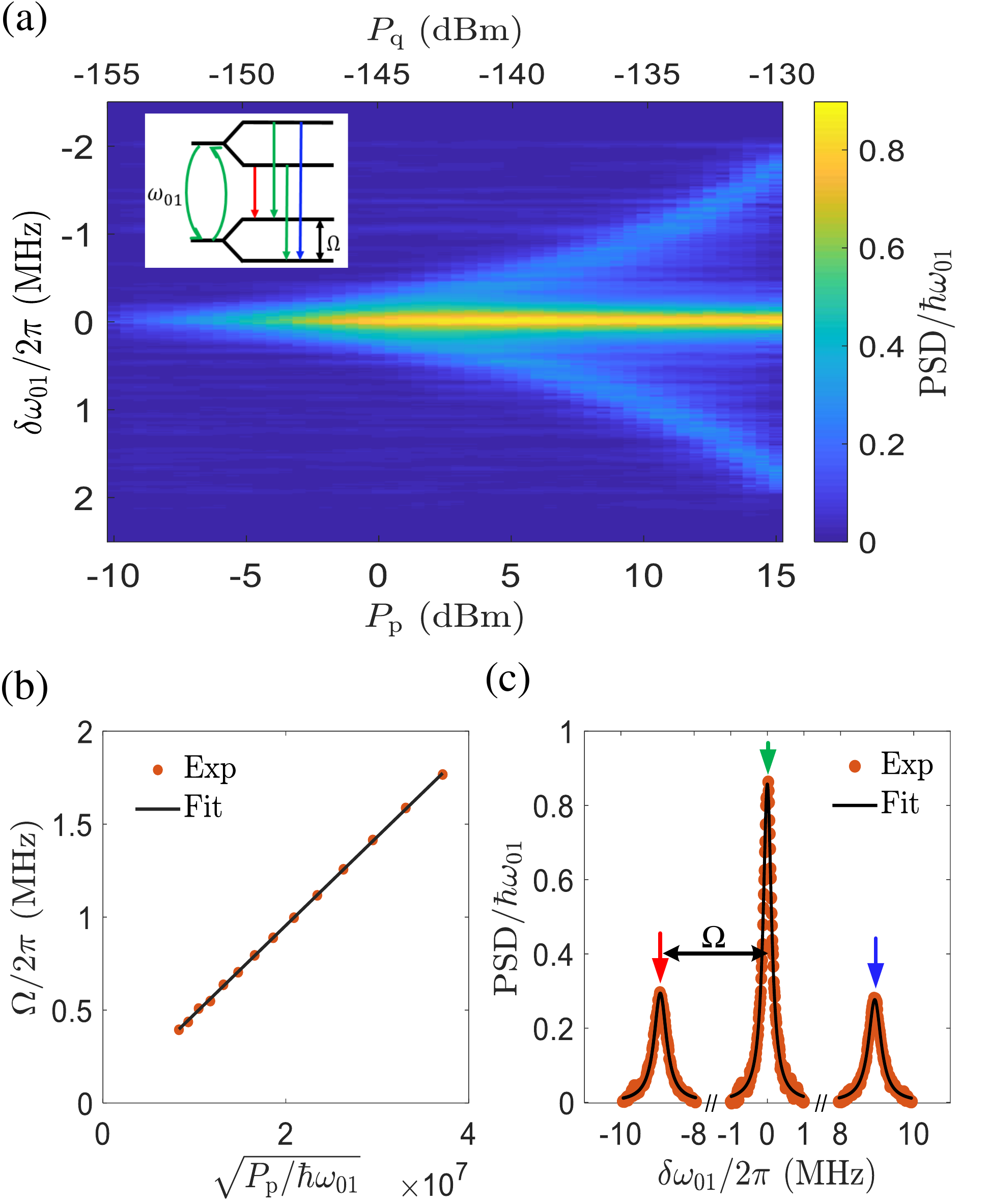}
\caption{Power spectral density (PSD) of the on-resonant Mollow triplet from the atomic fluorescence.
(a) Resonant fluorescence emission spectrum as a function of the pump power and detuning of the detected radiation, $\delta\omega_{01}=\omega-\omega_{01}$. $P_{\rm{p}}$ is the power from the RF source while $P_{\rm{q}}$ is the corresponding power on the qubit. Inset: a schematic of the triplet transitions in the dressed-state picture,  where the qubit energy levels split by $\Omega$ due to strong driving, creating three transitions with frequencies $\omega_{01}-\Omega$, $\omega_{01}$ and $\omega_{01}+\Omega$.
(b) Rabi splitting $\Omega$ (dots) vs drive amplitude, extracted from (a). The black line is the linear fit to obtain the attenuation in the input line which is A = -145dB.
(c) Power spectral density at -116\,\rm{dBm} power at the sample. The black lines are individual fits to the linewidths of the three peaks, yielding $\Gamma_2/2\pi=141\pm2\,\rm{kHz}$ and $\Gamma_1/2\pi=276\pm5\,\rm{kHz}$. The arrows correspond to the transitions in the inset of (a).
}
\label{onresonanceMollow}
\end{figure}

Under resonant continuous microwave excitation ($\Delta=\omega_{\rm{p}}-\omega_{01}=0$), as shown in Fig.~\ref{onresonanceMollow}(a), the splitting between the sidebands and the central peak increases as the pump power $P_{\rm{p}}$. The splitting equals the Rabi frequency and obeys $\Omega=2\sqrt{A\Gamma_rP_{\rm{p}}/(\hbar\omega_{01})}$. By fitting the extracted Rabi splitting $|\Omega|$ in Fig.~\ref{onresonanceMollow}(b), and using $\Gamma_r$ from the previous measurement in Sec.\,\uppercase\expandafter{\romannumeral2},  we extract a total attenuation $A=-145\,\rm{dB}$ of which about -125\,\rm{dB} attenuation is from attenuators and directional couplers, -7\,\rm{dB} from an Eccosorb filter and the rest is due to cable loss. This allows us to renormalize all applied powers to either the power at the qubit, or the corresponding Rabi frequency. The total gain in the output line of the measurement setup can be calibrated by tuning the qubit away and measuring the power at the output port at room temperature. This results in a total gain $G=115\,\rm{dB}$, of which approximately 44\,\rm{dB} gain comes from a high electron mobility transistor (HEMT) amplifier, and the rest is from the room temperature amplifiers and the pre-amplifiers of the digitizer.

The Rabi rate can be made much larger than all the decay rates of the qubit ($\Omega\gg\Gamma_1,\Gamma_2$). Consequently, the overlap in the frequency domain between the sideband emission and the central peak becomes negligible. In Fig.~\ref{onresonanceMollow}(c), we use an input power to the qubit $P_{\rm{q}}\approx-116\,\rm{dBm}$, equivalent to $\Omega/2\pi\approx 9\,\rm{MHz}$. The incoherent part of the corresponding power spectral density (PSD) is given by
\begin{eqnarray}
S_i(\omega)&\approx &\frac{1}{2\pi}\frac{\hbar\omega_{01}\Gamma_{\rm{r}}}{4}\Big\{\frac{\Gamma_{\rm{s}}}{(\delta\omega_{01} + \Omega)^2 + \Gamma_{\rm{s}}^2}\nonumber\\
&+&\frac{2\Gamma_2}{\delta\omega_{01}^2+\Gamma_2^2}+\frac{\Gamma_{\rm{s}}}{(\delta\omega_{01}-\Omega)^2+\Gamma_{\rm{s}}^2}\Big\},
\label{PSDonresonance}
\end{eqnarray}
[see Eq.~(\ref{A.13}) in Appendix~\ref{sec:Spectrum}], where the half width at half maximum of the central peak and the sidebands are $\Gamma_2$ and $\Gamma_{\rm{s}}=(\Gamma_1+\Gamma_2)/2$, respectively. The solid curves in Fig.~2(c) are fits to Eq.~(\ref{PSDonresonance}) using that the PSD expressed in linear frequency is $2 \pi S_i(\omega)$. We obtain $\Gamma_2/2\pi=141\pm2\,\rm{kHz}$ for the central peak, $\Gamma_{\rm{s, red}}/2\pi=210\pm3\,\rm{kHz}$ and $\Gamma_{\rm{s, blue}}/2\pi=206\pm4\,\rm{kHz}$ for sidebands. By taking the average of $\Gamma_{\rm{s,red}}$ and $\Gamma_{\rm{s,blue}}$, we obtain  $\Gamma_1/2\pi=275\pm7\,\rm{kHz}$. We note that the extracted $\Gamma_2/2\pi$ value is fully consistent with the result from the reflection-coefficient measurement in Sec.\,\uppercase\expandafter{\romannumeral2}. From that measurement, we also know  $\Gamma_{\rm{r}}/2\pi= 227\pm 1\,\rm{kHz}$. Thus, we can now extract both the non-radiative decay rate, $\Gamma_{\rm{n}}/2\pi=48 \pm7\,\rm{kHz}$, and the pure dephasing rate, $\Gamma_\phi/2\pi=3\pm4\,\rm{kHz}$.

By integrating the PSD of each peak in the Mollow triplet we can compare their relative weights. After normalization with $\hbar\omega_{01}\Gamma_{\rm{r}}$, the results are about 0.254, 0.116 and 0.124 for the middle peak, the red and blue sidebands, respectively. According to Eq.~(\ref{PSDonresonance}), we would expect these numbers to be 0.250, 0.125 and 0.125, respectively, for a fully saturated qubit. 

\subsection{Off-resonant Mollow triplet}

\begin{figure}[H]
\includegraphics[width=\linewidth]{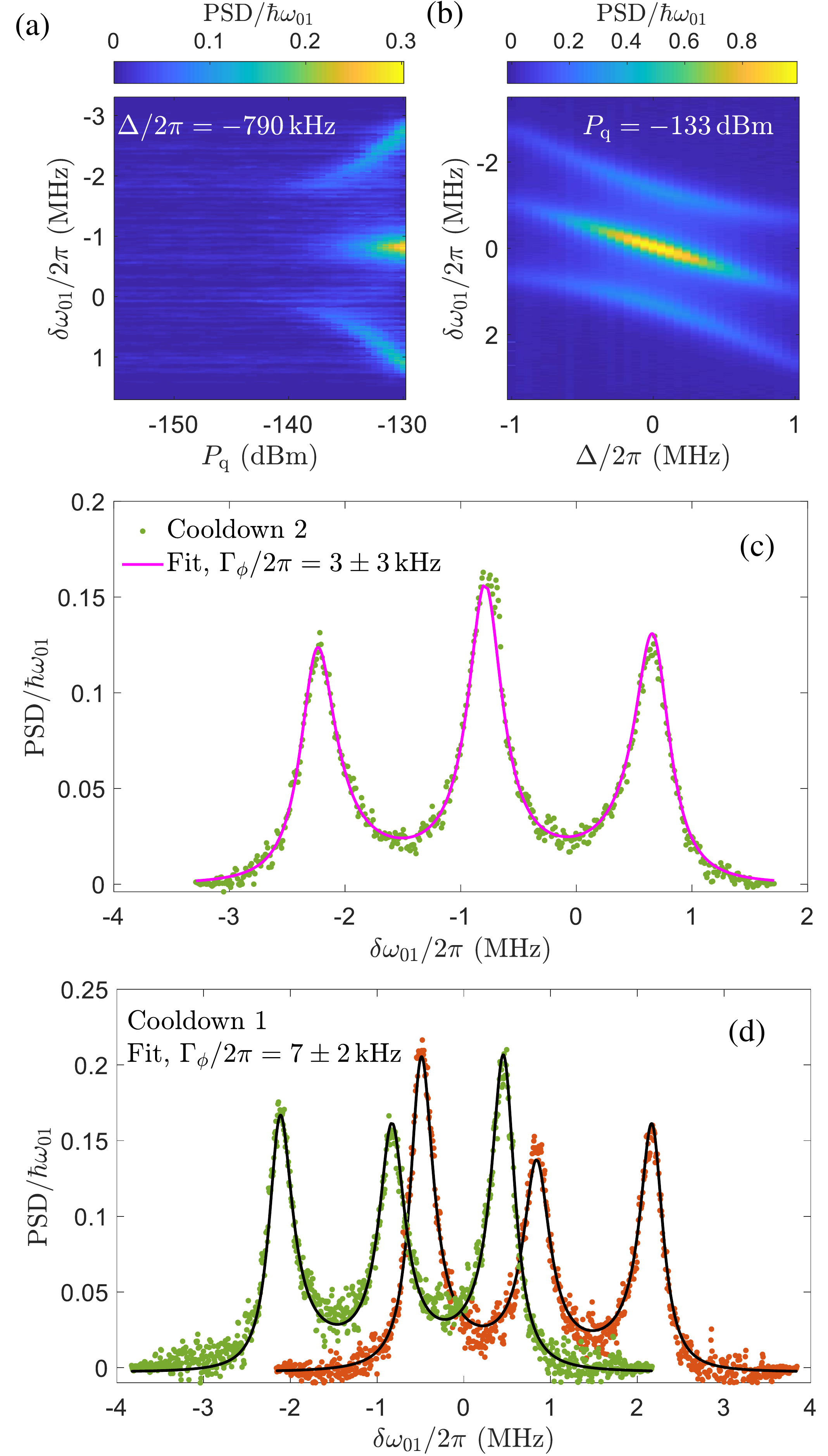}
\caption{Power spectral density of the off-resonant Mollow triplet from the atomic fluorescence.
(a) Off-resonant fluorescence emission spectrum as a function of the pump power at the qubit. The frequency of the pump field is detuned by $\Delta=\omega_{\rm{p}}-\omega_{01}=-2\pi*790\,\rm{kHz}$ from the qubit frequency. The Rabi splitting is increased with the input power.
(b) Off-resonant fluorescence emission spectrum as a function of the frequency of the pump at $P_{\rm{q}}=-133\,\rm{dBm}$.
(c) Off-resonant PSD at $\Delta/2\pi=-790\,\rm{kHz}$ (green dots) with $P_{\rm{q}}=-133\,\rm{dBm}$. The solid curve is a fit to $\Gamma_1/2\pi=275\pm6\,\rm{kHz}$ and $\Gamma_{\phi}/2\pi=3\pm3\,\rm{kHz}$.
(d) Off-resonant PSD in a second cooldown at $\Delta/2\pi=820\,\rm{kHz}$ (brown dots), $\Delta/2\pi=-830\,\rm{kHz}$ (green dots). A value of $\Gamma_{\phi}/2\pi=7\,\rm{kHz}$ gives a good fit to both traces (black).
}
\label{offresonanceMollow}
\end{figure}

We also study the off-resonant Mollow triplet at a variety of pump powers and frequency detunings between the pump and the qubit.
In Fig.~\ref{offresonanceMollow}(a), the pump power at the qubit is swept from -150\,dBm to -130\,dBm at detuning $\Delta/2\pi=790\,\rm{kHz}$. We find that the PSD is weaker than in the on-resonant case, implying that the qubit is less excited. In Fig.~\ref{offresonanceMollow}(b), as we sweep the frequency detuning between the pump and the qubit, the spectrum over 1\,\rm{MHz} bandwidth is nearly symmetric, so that the extracted pure dephasing rates by the off-resonant fluorescence are insensitive to the frequency detuning $\Delta$. We can either choose a large $\Delta$ which will lead to a small excitation of the qubit, or a small $\Delta$  which results in an unresolved spectrum between the central peak and sidebands.

Compared to the on-resonant case, the off-resonant Mollow triplet carries additional information in its sideband asymmetry and the approximation used in Eq.~(\ref{PSDonresonance}) is no longer valid. Therefore, the full expression for the PSD must be used, shown in Eq.~(\ref{A.12}) in Appendix~\ref{sec:Spectrum} which is an extension of Ref.~\cite{koshino2012control}. The fit of the data in Fig.~\ref{offresonanceMollow}(c) to Eq.~(\ref{A.12}) yields $\Gamma_\phi/2\pi=3\pm3\,\rm{kHz}$ (pink solid line). The symmetry of the sidebands around the central peak is due to the relatively small pure dephasing rate. In Fig.~\ref{offresonanceMollow}(d), the sample was measured in an earlier cooldown. There, we observed a larger asymmetry for both positive (green dots) and negative detunings (brown dots). In the case of positive detuning, the red sideband is closer to the qubit original frequency than the blue sideband, whereas the blue sideband is closer when the sign of the detuning is changed. We fit the two set of data simultaneously to obtain $2\pi*(7\pm2)\,\rm{kHz}$, which is slightly larger than the second cooldown. This is likely due to that we used only two isolators in the first cooldown, and four isolators in the second cooldown, leading to less thermal photons from the transmission line in the second case.

\begin{table}
 \caption{Summary of the different rates extracted using different methods. The first method is a measurement of the reflection coefficient under a weak probe [Sec.\,\uppercase\expandafter{\romannumeral2}]. On-res.MT and Off-res.MT represent the on/off-resonant Mollow triplet from atomic fluorescence, respectively [Sec.\,\uppercase\expandafter{\romannumeral3}]. The energy loss is estimated by calibrating the input power and measuring both the coherent and incoherent power scattered by the atom [Sec.\,\uppercase\expandafter{\romannumeral4}]. Finally, time resolved measurements of the decay from both a superposition state and the first excited state were used [Sec.\,\uppercase\expandafter{\romannumeral5}]. BW and T are the measurement bandwidth and the total measurement time, respectively, for each method.} \label{tab:2}
  \centering
\begin{tabular*}{\columnwidth}{  @{\extracolsep{\fill}} c c c c c c c c  @{} }
  \hline
  \hline
  Method &$\Gamma_{\rm{r}}/2\pi$ & $\Gamma_{\rm{n}}/2\pi$ & $\Gamma_{\phi}/2\pi$ &$\Gamma_1/2\pi$&$\Gamma_2/2\pi$&BW&T\\
         &kHz&kHz&kHz&kHz&kHz&MHz&h\\
  \hline
  Reflection & 227 (1)& - & - & - & 141 (1) & 4 & 3\\
  On-res.MT & - & 48 (7) & 3 (4) & 275 (7) & 141 (2) & 2 & 8\\
  Off-res.MT & - & 48 (6) & 3 (3) & 275 (6) & 140 (3) & 5 & 23\\
  Scattering & 229 (2) & 49 (1) & 1 (1) & 278 (2) & 140 (1) & 5 & 63\\
  SinglePoint & - & 50 (3) & 2 (2) & 277 (2) & - & 5 & 2\\
  Dynamics & - & 46 (11) & 9 (5) & 273 (11) & 145 (1) & 20 & 23\\
  \hline
  \hline
\end{tabular*}

\end{table}

\begin{figure}[]
\includegraphics[width=\linewidth]{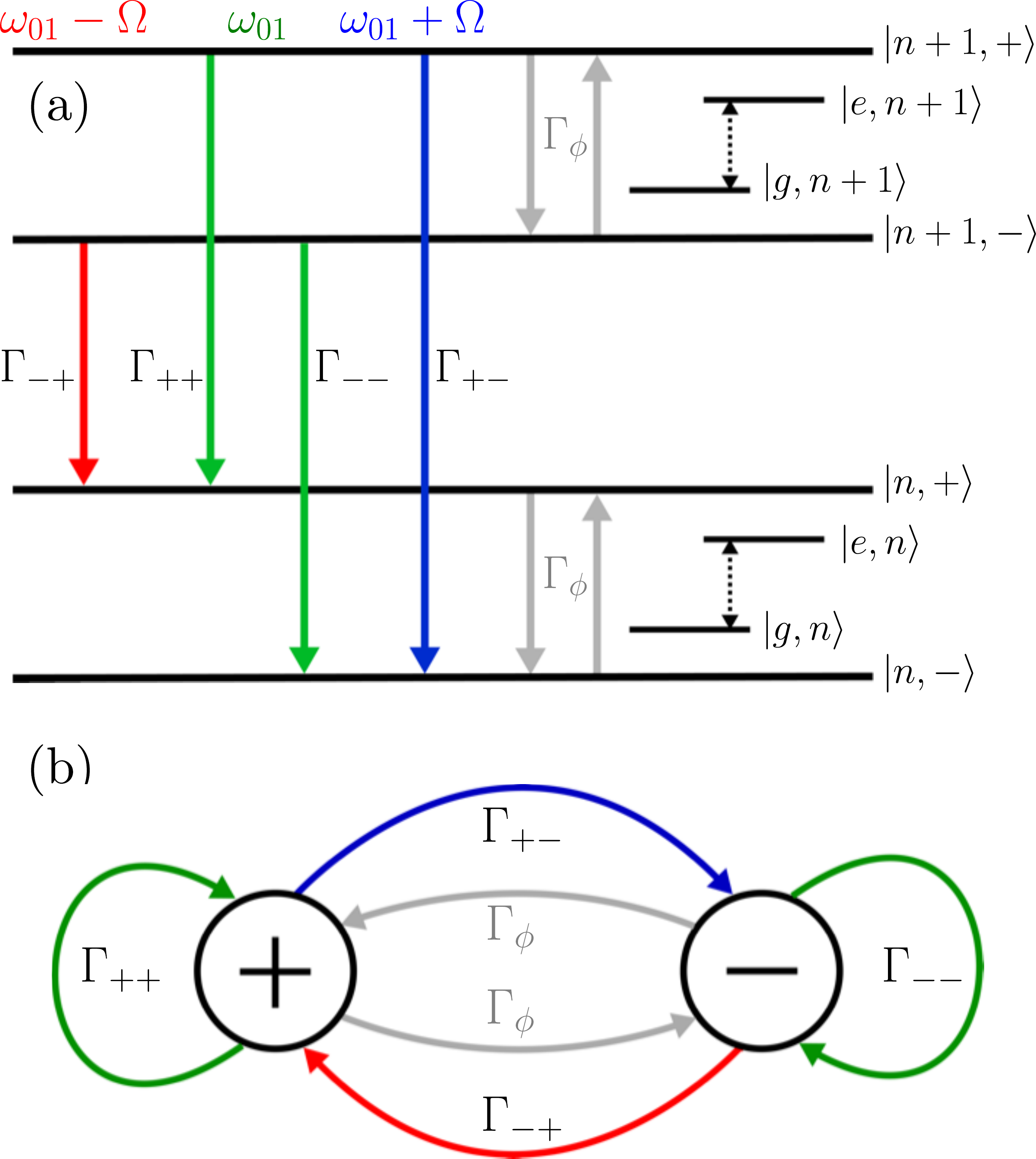}
\caption{Dressed states and transitions of the driven qubit.
(a) Sketch of the dressed-state picture, including energies and transition rates. The states $\ket{e, n}$, $\ket{g, n}$, $\ket{e, n+1}$, and $\ket{g, n+1}$ are the bare states;
the states $\ket{n,\pm}$ and $\ket{n+1, \pm}$ are the dressed states.
(b) Transitions and transition rates between the $+$ and $-$ subspaces.
}
\label{offresonatTheory}
\end{figure}

The mechanism by which the pure dephasing gives rise to an asymmetry in the Mollow triplet can be understood as follows (for details, see Appendix~\ref{sec:Asymmetric}). Relaxation from the qubit will cause transitions between dressed states $\ket{n, \pm}$ [see Fig.~\ref{offresonatTheory}(a)] that contain different numbers $n$ of drive photons. As shown in Fig.~\ref{offresonatTheory}(b), these transitions will either be between or within the $+$ and $-$ subspaces. In equilibrium, if the pure dephasing rate is zero, the probabilities $P_\pm$ for the system to be in these subspaces are given by the detailed-balance condition
\begin{equation}
\Gamma_{+-} P_+ = \Gamma_{-+} P_-,
\end{equation}
i.e., the number of emitted photons causing transitions from $+$ to $-$ (the blue sideband) must equal the number of emitted photons causing transitions from $-$ to $+$ (the red sideband). However, the interaction causing pure dephasing has a non-zero matrix element for transitions between $\ket{n, +}$ and $\ket{n, -}$, which leads to a modified detailed-balance condition:
\begin{equation}
(\Gamma_{+-} + \Gamma_\phi) P_+ = (\Gamma_{-+} + \Gamma_\phi) P_-.
\end{equation}
As $\Gamma_\phi$ increases, this will push the occupation probabilities towards $P_+ = P_-$. For off-resonant driving $\Gamma_{+-} \neq \Gamma_{-+}$ and thus the number of emitted photons in the two sidebands becomes different: $\Gamma_{+-} P_+ \neq \Gamma_{-+} P_-$. The larger number of photons will be emitted at the frequency corresponding to the larger of the two transition rates $\Gamma_{+-}$ and $\Gamma_{-+}$; from transition-matrix elements, this can be seen to be the frequency that is closer to the qubit frequency.

From Fig.~\ref{offresonanceMollow}(c), we have $\Gamma_1/2\pi=275\pm6\,\rm{kHz}$ and $\Gamma_2/2\pi=140\pm3\,\rm{kHz}$. Again, from the measured radiative decay rate, by subtracting $\Gamma_{\rm{r}}$ from $\Gamma_1$, we obtain $\Gamma_{\rm{n}}/2\pi = 48 \pm 6~\rm{kHz}$. Based on the results in this section, the on/off-resonant Mollow spectra allow us to extract the pure dephasing rate and non-radiative rate of a qubit. Specifically, for our qubit in this environment, we find that the non-radiative decay rate is one order of magnitude larger than the pure dephasing rate. Compared to the on-resonant Mollow triplet, the off-resonant Mollow triplet allows us to characterize the qubit decay rates at a lower pump power.

\section{Photon scattering by the qubit}
\label{sec3}
\begin{figure}[]
\includegraphics[width=\linewidth]{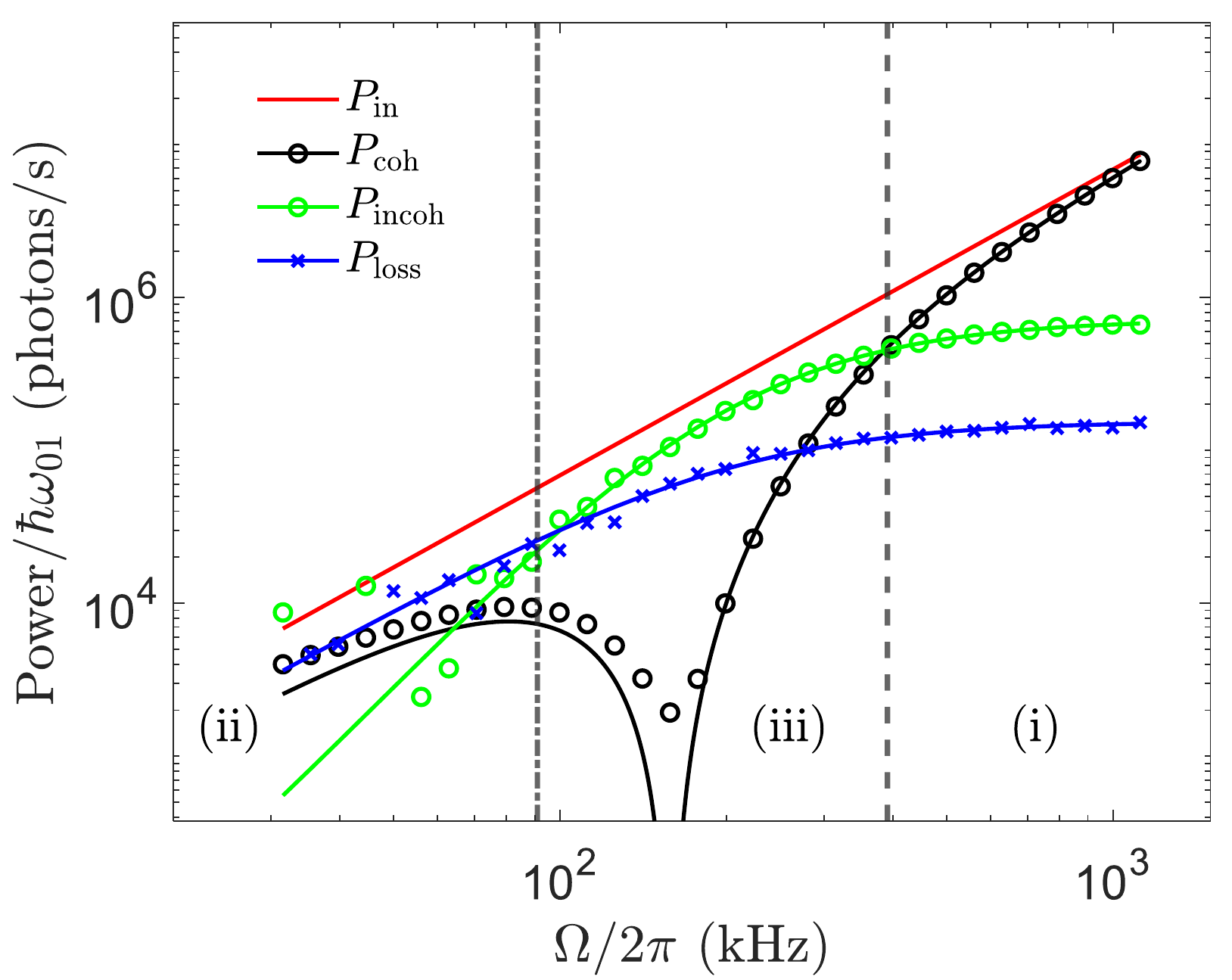}
\caption{Normalized powers as a function of Rabi frequency.
The input power, $P_{\rm{in}}$ (red), representing the input photon flux at the qubit, is measured when the qubit is tuned away by the external flux. When the qubit is on resonance with the input signal, we have the coherent power, $P_{\rm{coh}}$ (black circles), consisting of photons reflected from either the qubit or the end of the transmission line. The qubit can also scatter photons incoherently, $P_{\rm{incoh}}$ (green circles), due to the decoherence of the qubit. Moreover, the excited qubit has some probability to release a photon to the environment resulting in the power loss, $P_{\rm{loss}}$ (blue crosses). The \mbox{solid} curves are fits to different types of the scattered powers. The dotted and dash-dotted lines separate the qubit response into three interesting regions (see more details in the text).}
\label{outputpower}
\end{figure}
To verify the extracted decay rates above, we can also \mbox{measure} the power scattered by the qubit and the \mbox{dissipated} power due to the non-radiative decay channel directly. We normalize all the powers by the single-photon energy $\hbar\omega_{01}$. The pump is on resonance with the qubit. The output power then consists of a coherent part and an incoherent part, $P_{\rm{out}}=P_{\rm{coh}}+P_{\rm{incoh}}$, where
$P_{\rm{coh}}=\frac{\Omega^2}{4\Gamma_{\rm{r}}}(1-\frac{\Gamma_1\Gamma_{\rm{r}}}{\Omega^2+\Gamma_2\Gamma_{\rm{1}}})^2$ and $P_{\rm{incoh}} =\frac{\Gamma_{\rm{r}}}{2}\frac{\Omega^2(\Gamma_1\Gamma_\phi+\Omega^2)}{(\Gamma_1\Gamma_2+\Omega^2)^2}$ (see Appendix~\ref{sec:Power}).

For our qubit, the pure dephasing rate was verified to be around $3\,\rm{kHz}$, i.e., much less than other rates and therefore negligible, so, the expression for the incoherent power can be further simplified to $P_{\rm{incoh}}\simeq 2\Gamma_{\rm{r}}\rho_{11}^2$, where $\rho_{11}$ is the population of the first excited state of the qubit. In this cover, 
the expression for the dissipated power due to the non-radiative decay is then $P_{\rm{loss}}=\Gamma_{\rm{n}}\rho_{11} = \Gamma_{\rm{n}}\frac{\Omega^2}{2(\Gamma_1\Gamma_2+\Omega^2)}$.

Experimentally, we use about $4.2\times10^9$ averages to measure all the powers. We denote the measured voltage $V$, the system noise $N$, and the pump power $P_{\rm{in}}$. The subscripts ``off'' and ``on'' used in the following contexts mean that the qubit is off/on resonance with the pump, respectively. When the qubit is tuned away, it is off-resonant with the pump; we will have $P_{\rm{in}}=\langle V\rangle^2_{\rm{off}}$ because of the coherence of the pump. Besides the pump power, the system noise will also make a contribution to the total measured power, $P_{\rm{in}}^{\rm{meas.}}$. Therefore, we have  $P_{\rm{in}}^{\rm{meas.}}=\langle V^2\rangle_{\rm{off}}=P_{\rm{in}}+N$.
When instead the qubit is on resonance with the pump, the total measured output power $P_{\rm{out}}^{\rm{meas.}}=\langle V^2\rangle_{\rm{on}}=P_{\rm{out}}+N$, where $P_{\rm{coh}}=\langle V\rangle^2_{\rm{on}}$ (black circles). Therefore, $P_{\rm{loss}}$ (blue crosses) is obtained by taking $P_{\rm{loss}}=P_{\rm{in}}^{\rm{meas.}}-P_{\rm{out}}^{\rm{meas.}}=P_{\rm{in}}-P_{\rm{out}}$ with $P_{\rm{incoh}}=P_{\rm{in}}-P_{\rm{coh}}-P_{\rm{loss}}$ (green circles). Figure~\ref{outputpower} shows all the types of measured power as a function of the Rabi frequency. There, we find three interesting regions:

\begin{enumerate}[label=(\roman*),labelindent=1em]
\item At high input power, when $\Omega>(1+\frac{1}{\sqrt{2}})\Gamma_{\rm{r}}=2\pi*391\,\rm{kHz}$, to the right of the dashed line in Fig.~\ref{outputpower}, the qubit starts to be saturated. The outgoing field is then mainly coherent from the pump itself. By increasing the input power further, the qubit is completely saturated, leading to $P_{\rm{in}}\approx P_{\rm{coh}}\approx\frac{\Omega^2}{4\Gamma_{\rm{r}}}$, $P_{\rm{incoh}}\approx\frac{\Gamma_{\rm{r}}}{2}$ and $P_{\rm{loss}}\approx\frac{\Gamma_{\rm{n}}}{2}$. In this case, almost all the incoming photons are reflected by the mirror.
\item In the low-power region ($\Omega<\sqrt{\frac{\Gamma_1\Gamma_2\Gamma_{\rm{n}}}{\Gamma_{\rm{r}}}}=2\pi*91~\rm{kHz}$ derived from $P_{\rm{incoh}}=P_{\rm{loss}}$,
 to the left of the dash-dotted line), the scattering process is dominated by the interaction between the qubit and the incoming photons. The incoherent scattering is proportional to $\rho_{11}^2$ whereas the power loss depends linearly on the excitation probability. Therefore, the incoherent power can be less than the power loss when $\rho_{11} < \frac{\Gamma_{\rm{n}}}{2\Gamma_{\rm{r}}}\approx0.11$. Besides the incoherent photons, there is a small coherent scattering by the qubit. Compared to the loss, the coherent power is smaller if the non-radiative decay is large enough, namely if $\Gamma_{\rm{n}}>\frac{\Gamma_1(\Gamma_{\rm{r}} - \Gamma_2)^2}{2\Gamma_{\rm{r}}\Gamma_2}\approx 2\pi*34\,\rm{kHz}$.
\item In the intermediate-power region where $\sqrt{\frac{\Gamma_1\Gamma_2\Gamma_{\rm{n}}}{\Gamma_{\rm{r}}}}<\Omega<(1+\frac{1}{\sqrt{2}})\Gamma_{\rm{r}}$, both the mirror and the qubit make substantial contributions to the scattering process. The photons reflected by the mirror interfere destructively with those scattered by the qubit, resulting in a suppression of the coherent part of the output field. In particular, the dip around $\Omega/2\pi\approx160~\rm{kHz}$ in the coherent power appears due to the fully destructive interference. In addition, the qubit excitation is not small anymore and the incoherent power is larger than the loss because $\Gamma_r>\Gamma_n$. We note that this region can be non-existent when either the non-radiative decay or the pure dephasing is sufficiently large.
\end{enumerate}

We also fit the data in Fig.~\ref{outputpower} to obtain all the decay rates. The result for the incoherent power indicates $\Gamma_{\rm{r}}/2\pi=229\pm2\,\rm{kHz}$ with $\Gamma_1\Gamma_2/4\pi^2=39590\pm211\,\rm{kHz^2}$ and $\Gamma_1\Gamma_\phi/4\pi^2=281\pm 281\,\rm{kHz^2}$. From fits to the power loss, we find $\Gamma_{\rm{n}}/2\pi=49\pm 1\,\rm{kHz}$ and $\Gamma_1\Gamma_2/4\pi^2=41260\pm4750\,\rm{kHz^2}$. Therefore, with $\Gamma_{\rm{r}}$ and $\Gamma_{\rm{n}}$, we obtain $\Gamma_1/2\pi=(\Gamma_{\rm{n}} + \Gamma_{\rm{r}})/2\pi=278\pm2\,\rm{kHz}$. Then, $\Gamma_\phi/2\pi\simeq1\,\rm{kHz}$.  The coherent power yields $\Gamma_{\rm{r}}/2\pi=229\pm2\,\rm{kHz}$, $\Gamma_{\rm{n}}/2\pi=48\pm8\,\rm{kHz}$ and $\Gamma_\phi/2\pi=1\pm1\,\rm{kHz}$. Then, with $\Gamma_1$ and $\Gamma_\phi$, we have $\Gamma_2/2\pi=140\pm1\,\rm{kHz}$.

From the discussion on region (i), at the highest Rabi frequency $\Omega/2\pi=1119$\,kHz in Fig.~\ref{outputpower}, $\Gamma_n/\Gamma_r\approx P_{\rm{loss}}/P_{\rm{incoh}}$. Then, we obtain $\Gamma_n/\Gamma_r=[0.1971, 0.2385, 0.2227, 0.2297]$, by dividing the total measured data into four pieces. Combined with $\Gamma_r$ from the reflection coefficient in Sec.\,\uppercase\expandafter{\romannumeral2}, we find $\Gamma_n/2\pi\approx 45, 54, 51, 52$\,kHz, respectively. The mean value is about 50\,kHz with 3\,kHz as the standard deviation. According to Eqs.~(\ref{C.4}) and (\ref{C.5}) in Appendix~\ref{sec:Power}, as $\Gamma_{\phi}$ is small for our qubit, we have $\Gamma_n=2P_{\rm{loss}}(1+\frac{\Gamma_1\Gamma_2}{\Omega^2})$ and $\Gamma_r\simeq2P_{\rm{incoh}}(1+\frac{\Gamma_1\Gamma_2}{\Omega^2})^2$. Due to $\frac{\Gamma_1\Gamma_2}{\Omega^2}\approx 3\%$, $\Gamma_n\approx2.06P_{\rm{loss}}$ and $\Gamma_r\approx2.12P_{\rm{incoh}}$. Therefore,  the estimated value for $\Gamma_n$ has a systematic error of about $3\%$. Since $\Gamma_{\phi}=\Gamma_2-\frac{\Gamma_r+\Gamma_n}{2}$ and $\Gamma_1=\Gamma_r+\Gamma_n$, the pure dephasing and the non-radiative decay rates are $2\pi*(2\pm2)$\,kHz and $2\pi*(277\pm2)$\,kHz, respectively. Additionally, because $P_{\rm{loss}}=2\pi*0.0243*10^6$ and $P_{\rm{incoh}}=2\pi*0.1056*10^6$, we have $\Gamma_n/2\pi\approx49\,\rm{kHz}$ and $\Gamma_r/2\pi\approx224\,\rm{kHz}$, respectively.



The results shown here agree well with the values from other sections in the paper, which implies that it is possible to take the $\Gamma_{\rm{r}}$ value from the reflection coefficient measurement as a reference for the Mollow triplet in order to separate the non-radiative decay rate and the pure dephasing rate.
\begin{figure}[tbph]
\includegraphics[width=\linewidth]{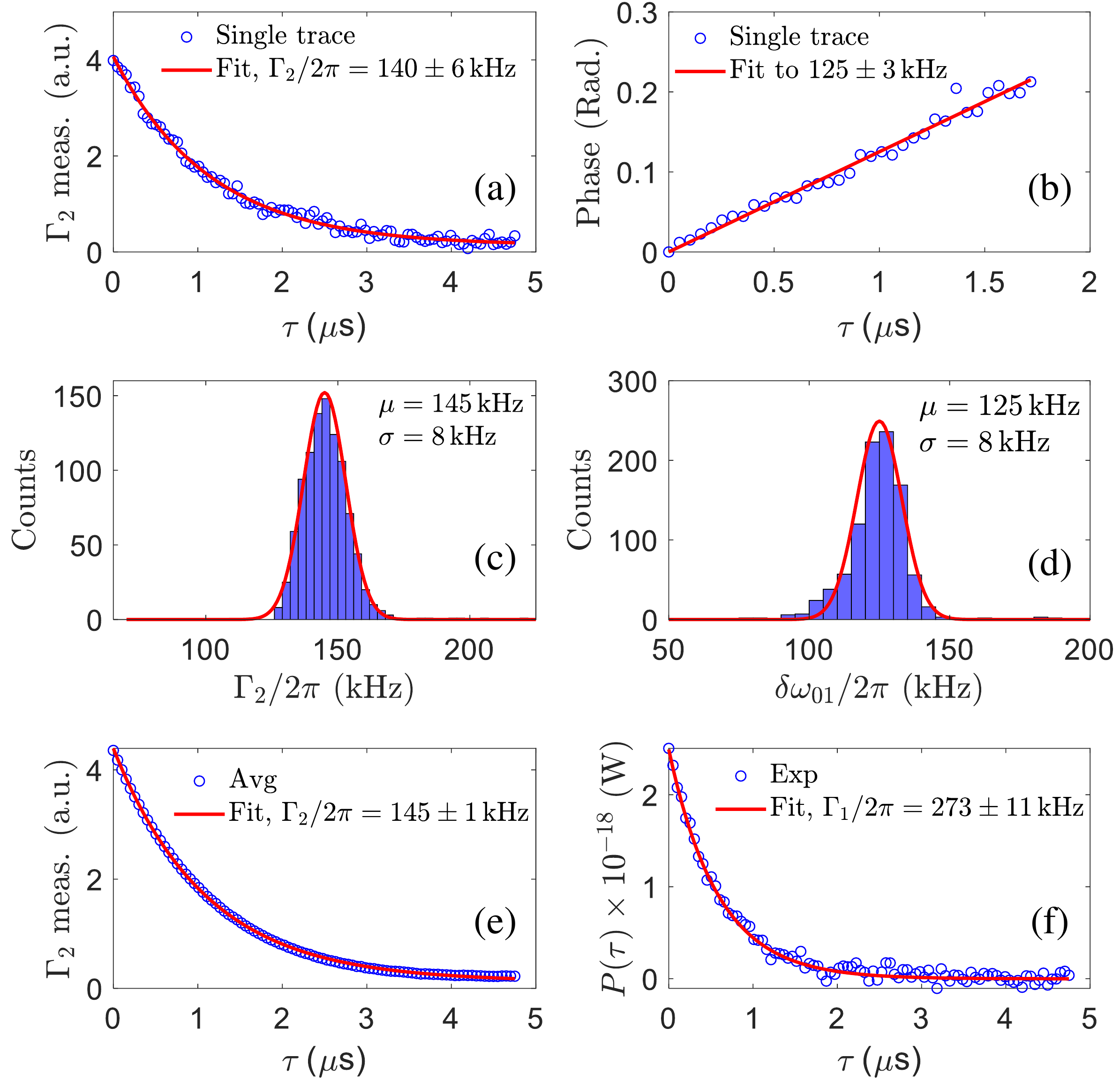}
\caption{Qubit dynamics to measure $\Gamma_2$, $\Gamma_1$ and $\delta\omega_{01}/2\pi$. The magnitude response of the measured signal is proportional to the magnitude of the emission operator $\langle\sigma_-\rangle$ of the qubit while the phase response increases linearly with time as $\delta\omega_{01} \tau$.
(a) A single trace of the magnitude response of $\Gamma_2$ measurements after a $\pi/2$-pulse, showing the decoherence processes of the qubit within time $\tau$. The data is fitted to an exponential decay.
(b) The corresponding phase response from (a), showing that the phase of the emitted photon from the qubit evolves with a slope corresponding to the detuning $\delta \omega_{01}/2\pi=125\,\rm{kHz}$ where $\delta\omega_{01}=\omega_{01}-\omega_{\rm{pulse}}$.
(c) Histogram of $\Gamma_2$ from the magnitude response of the measurements from 975 traces, spanning $4.27\times10^5$ s (approximately 119 h).
(d) Histogram of $\Gamma_2$ from the corresponding phase response of the measurements taken in (c). Both (c) and (d) have been fitted (solid line) to a Gaussian distribution with parameters shown.
(e) The decay of the qubit state by averaging all the measured traces in (c) to extract the decoherence rate.
(f) A $\pi$-pulse is applied to flip the qubit to the excited state with $1.92\times10^9$ averages, where $P(\tau)$ is the power emitted by the qubit at time $\tau$ after the pulse.  By fitting the emitted power (blue circles) to an exponential decay, we extract $\Gamma_1/2\pi=273\pm 11\,\rm{kHz}$. Except for the histograms, the error bars are for 95\% confidence.
}
\label{T2Mag}
\end{figure}
\section{Time-resolved dynamics}
\label{sec4}


All measurements described previously span time ranges from several hours to tens of hours. It is noteworthy that qubit decay rates extracted by different methods agree relatively well. However, the long duration means that any fluctuations of the decay rates are averaged out. Recently, several groups have characterized such fluctuations in circuit QED, using Rabi pulses, Ramsey interference measurements, and dispersive qubit readout~\cite{muller2015interacting,burnett2019decoherence,schlor2019correlating,klimov2018fluctuations}. To probe the decoherence of the qubit with a temporal resolution of 7 minutes, we prepare the qubit in a superposition of the ground and first-excited state and monitor its spontaneous emission into the waveguide by recording both quadratures of the output field with a digitizer as a complex trace in the time domain. We measure for $4.27\times10^5$ s (approximately 119 hours) with 975 repetitions, and each such trace has $2.30\times10^6$ averages.

After the 50 ns long $\pi/2$-pulse, the qubit superposition state evolves in time $\tau$ as $\frac{1}{\sqrt{2}}(|0\rangle+e^{-\Gamma_2\tau-i\delta \omega_{01} \tau}|1\rangle)$ with $\delta \omega_{01}=\omega_{01}-\omega_{\rm{pulse}}$. The emitted field carries information about the qubit operator $\langle\sigma_-\rangle=e^{-\Gamma_2\tau} e^{-i\delta \omega_{01}\tau}$, where the amplitude response and the phase response show the decoherence and the qubit frequency shift with $\tau$, respectively.

Figures~\ref{T2Mag}(a) and (b) show the magnitude and phase response of a single trace where the decay of the magnitude is fitted to an exponential curve and  the phase of the photons emitted from the qubit grows linearly with time due to the free evolution of the qubit, where the slope determines $\delta \omega_{01}/2\pi=125\,\rm{kHz}$. Figures~\ref{T2Mag}(c) and (d) are histograms of $\Gamma_2$ and $\delta\omega_{01}$ for all the repetitions. Both histograms can be fitted to a Gaussian with parameters shown in the figures. In comparison with the decoherence rates extracted from other measurements, we find that the standard deviation here is larger than the previously measured error bar. This shows that the dynamics of the qubit on a short time differs slightly from that over a long measurement time. By taking the average of all the traces in Fig.~\ref{T2Mag}(d), we fit to an exponential decay and get an averaged $\Gamma_2/2\pi=145\pm1\,\rm{kHz}$.

To also study $\Gamma_1$, we instead send a $\pi$-pulse to flip the qubit fully, and then measure the emission from the qubit. The corresponding output power, $P(\tau) = (\hbar\omega_{01}\Gamma_{\rm{r}}/2) (1 + \langle\sigma_z\rangle) e^{-\Gamma_1\tau}$~\cite{abdumalikov2011dynamics} allows us to determine $\Gamma_1$. The trace is measured with $1.92\times10^9$ averages, shown in Fig.~\ref{T2Mag}(e).  A fit to an exponential decay with $\Gamma_1/2\pi = 273\pm11\,\rm{kHz}$ agrees well with the data. Combining these numbers with $\Gamma_r$ from Sec.\,\uppercase\expandafter{\romannumeral2}, we can also calculate $\Gamma_n$ and $\Gamma_{\phi}$ from these measurements. The resulting values can be seen in Table~\ref{tab:2}.

\section{Discussion and conclusion}
\label{sec5}

We have shown several methods to determine different decay rates of a qubit placed in front of a mirror. In principle, these methods can also be used when the qubit couples to a transmission line without a mirror, except for the scattering method, where the corresponding measurement taken on both the input and output ports is required.

In our case, the measured rates are consistent between methods within the error bars of two standard deviation corresponding to $95\%$ confidence. The results are summarized in Table~\ref{tab:2}. The reflection measurement is the baseline to provide the value of $\Gamma_r$ to extract the non-radiative decay rate of the qubit for measurements except for the scattering measurement. These different methods have advantages and disadvantages that we summarize below:

\begin{enumerate}[label=(\roman*),labelindent=1em]
\item The fastest way to obtain the non-radiative decay rate is to send a strong pump on resonance with the qubit so that the central peak and the sidebands of the Mollow triplet do not overlap. The drawback is that the pump power needed here is much stronger than for the other methods and that may change the rates slightly.
\item In the second method, we measure the off-resonant Mollow triplet by detuning the pump frequency slightly from the qubit frequency. The sidebands will be asymmetric around the central peak if the pure dephasing rate is non-negligible. In this case, only weak probe power is required. However, the corresponding measurement time is increased by almost a factor of three.
\item The most accurate way to measure the non-radiative decay rate is to measure the difference between the input and output power, labelled as Scattering in Table~\ref{tab:2}. Using this method, we can obtain not only the power loss but also the coherent and incoherent power scattered by the qubit. However, the measurement time is much longer. In addition, the attenuation between the sample and the input line as well as  the gain between the detector and the sample need to be calibrated at the beginning in order to get the absolute power values from the qubit.  To simplify the measurement, as we discussed in Sec.\,\uppercase\expandafter{\romannumeral3}, we can use that when the pump saturates the qubit, we get $P_{\rm{incoh}}\approx\frac{\Gamma_{\rm{r}}}{2}$ and $P_{\rm{loss}}\approx\frac{\Gamma_{\rm{n}}}{2}$. Then, the ratio of the non-radiative decay rate to the radiative decay rate can be obtained from the ratio of the lost power to the incoherent power. Knowing the value of $\Gamma_{\rm{r}}$ from the reflection measurement, we can obtain the non-radiative decay rate. Therefore, in principle, we do not need to sweep the pump power as was done in Fig.~\ref{outputpower}. This simple way is labeled as SinglePoint in Table~\ref{tab:2}.
\item Finally, pulses can be applied to excite the qubit. Afterwards, the exponential decay of the emission and the emitted power trace from the qubit can be recorded to extract the total relaxation rate and the decoherence rate with a much larger measurement bandwidth. The distortion on the scattered photons due to the non-flat frequency response will affect the extracted values of the decay rates. This may be the reason why decay rates from this method are slightly different from those measured by other methods. However, the advantage of this method is that it allows us to study the short-time dynamics of the qubit.
\end{enumerate}

The measurement time for these methods are from 2 hours to 63 hours. The coherent measurement is related to the first moment (amplitude) whereas other methods are related to the second moment (power). 
In order to estimate the system noise $N$, we measured the background PSD by turning the drive off (not shown) and comparing the result with the measurement of Fig.~\ref{onresonanceMollow}(c). We found $N \approx 49$ photons.
 However, we expect that using a quantum-limited Josephson traveling-wave amplifier~\cite{macklin2015near} would reduce the system noise to about two photons.
 This would result in a reduction of the measurement time by factors of 5 and 25, respectively, for the coherent measurement and the other methods.

From the measured result, our qubit is $T_1$-limited, i.e., the radiative decay dominates the interaction. However, the non-radiative decay rate is one order of magnitude larger than the pure dephasing rate. The corresponding spontaneous-emission factor is $\beta\approx 85\%$, which is typically close to $100\%$ when we engineer the radiative decay much larger than the non-radiative decay. Therefore, to reduce the non-radiative decay rate will be the next step to improve the intrinsic coherence of our qubit. In addition, it is worthwhile to investigate why the non-radiative decay rate of our qubit is one order of magnitude larger than the qubit coupled to a resonator which was fabricated on the same wafer~\cite{burnett2019decoherence} in the future.

Our methods allow us to analyze all the decay channels in detail. This will be useful to study and engineer decay channels of the qubit, which is the critical element in superconducting circuits. For example, engineering the decay channels can improve the quantum efficiency of generating single photons, which is set by $\Gamma_r/2\Gamma_2$. Also, the fidelity of detecting a single photon can be increased by extending the qubit coherence time. More importantly, compared to circuit QED where a resonator couples to a qubit dispersively, our study provides a straightforward way to investigate superconducting qubits, which are crucial elements in superconducting quantum computers.

\section{ACKNOWLEDGMENTS}
We wish to express our gratitude to David Niepce and Marco Scigliuzzo for insightful discussions. We acknowledge financial support from the Knut, Alice Wallenberg Foundation, and the Swedish Research Council, and the EU-project OpenSuperQ.
\appendix

\section{Power Spectrum}\label{sec:Spectrum}


Here, we follow the method in Ref.~\cite{koshino2012control} to calculate our circuit model. Our qubit Hamiltonian is ($\hbar=1$)
\begin{equation}
H = -\frac{\Delta}{2}\sigma_z + \frac{\Omega}{2}\sigma_x,
\label{A.1}
\end{equation}
where $\Delta = \omega_{\rm{p}}-\omega_{01}$; $\omega_{\rm{p}}$ and $\omega_{01}$ are the pump frequency and the qubit $|0\rangle \leftrightarrow |1\rangle$ transition frequency, respectively.

The Lindblad master equation, describing the qubit dynamics with decoherence included, is given by
\begin{equation}
\frac{d}{dt}\rho=\mathcal{L}\rho = - i[H,\rho ] + {\mathcal{L}_{\gamma}}\rho,
\label{A.2}
\end{equation}
where the Liouvilian $\mathcal{L}_{\gamma}$ is
\begin{equation}
\begin{split}
   \mathcal{L}_\gamma \rho & =  \Gamma_1D[\sigma_{-}]\rho+\frac{\Gamma_\phi}{2}D[\sigma_{z}]\rho,
\end{split}
\label{A.3}
\end{equation}
in which $D[c]\rho = c\rho c^\dag - \frac{1}{2}(c^\dag c\rho + \rho c^\dag c)$.

In the frame rotating with $\omega_{\rm{p}}$, the corresponding equations of motion for $s_1(t)\equiv \rho_{10}(t) = \left \langle \sigma_-(t) \right \rangle e^{i\omega_{\rm{p}}t}$ and $s_2(t)\equiv\rho_{11}(t) = \left \langle \sigma_+(t)\sigma_-(t) \right \rangle $ are obtained from Eq.~(\ref{A.2})
\begin{equation}
\frac{d}{dt}\begin{pmatrix}
s_1\\
s_1^*\\
s_2\\
\end{pmatrix}
=
M
\begin{pmatrix}
s_1\\
s_1^*\\
s_2\\
\end{pmatrix}
+B,
\label{A.4}
\end{equation}
where
\begin{equation}
M=
\begin{pmatrix}
i\Delta-\Gamma_2 & 0 & i\Omega\\
0 & -i\Delta-\Gamma_2& -i\Omega\\
i\Omega/2 & -i\Omega/2&-\Gamma_1
\end{pmatrix},
B=\begin{pmatrix}
-i\Omega/2\\
i\Omega/2\\
0\\
\end{pmatrix}.
\label{A.5}
\end{equation}
Here, $\Gamma_1$, $\Gamma_\phi$, and $\Gamma_2 = \frac{1}{2}\Gamma_{1} + \Gamma_\phi$ are the total relaxation rate of the qubit, the pure dephasing rate, and the decoherence rate, respectively.

The qubit reaches its stationary state for $t\gg\Gamma_{1,2}^{-1}$, The stationary values, $\overline s_1 = s_1(\infty)$ and $\overline s_2 = s_2(\infty)$, are

\begin{eqnarray}
\overline{s_1}&=& \frac{\Omega\Gamma_1(\Delta - i\Gamma_2)}{2(\Omega^2\Gamma_2 + \Gamma_1(\Delta^2 + \Gamma_2^2))}, \\
\overline{s_2}&=& \frac{\Omega^2\Gamma_2}{2(\Omega^2\Gamma_2 + \Gamma_1(\Delta^2 + \Gamma_2^2))}.
\label{A.6}
\end{eqnarray}

To determine the two-time correlation function of the atom, three quantities are defined:
\begin{eqnarray}
s_3(\tau) &=& \left \langle \sigma_+(t)\sigma_-(t+\tau) \right \rangle e^{i\omega_{\rm{p}}\tau}, \\
s_4(\tau) &=& \left \langle \sigma_+(t)\sigma_+(t+\tau) \right \rangle e^{-i\omega_{\rm{p}}(2t+\tau)}, \\
s_5(\tau) &=& \left \langle \sigma_+\sigma_+(t+\tau)\sigma_-(t+\tau) \right \rangle e^{-i\omega_{\rm{p}}\tau},
\end{eqnarray}
all of which are time-independent when stationary. From Eq.~(\ref{A.4}), we have equations of motion for these quantities as
\begin{equation}
\frac{d}{dt}\begin{pmatrix}
s_3\\
s_4\\
s_5\\
\end{pmatrix}
=
M
\begin{pmatrix}
s_3\\
s_4\\
s_5\\
\end{pmatrix}
+
B,
\label{A.7}
\end{equation}
with initial values $s_3(0) = \overline s_2$ and $s_4(0) = s_5(0) = 0$. In the $\tau\rightarrow\infty$ limit, the stationary values are $\overline s_3 = \left | s_1 \right |^2$,
$\overline s_4 = (\overline s_1^*)^2$, and $\overline s_5 = \overline s_1^*\overline s_2$. Using new variables, $\delta s_j(\tau) = s_j(\tau)-\overline s_j (j = 3,4,5)$, the above equations are rewritten as
\begin{equation}
\frac{d}{dt}\begin{pmatrix}
\delta s_3\\
\delta s_4\\
\delta s_5\\
\end{pmatrix}
=
M
\begin{pmatrix}
\delta s_3\\
\delta s_4\\
\delta s_5\\
\end{pmatrix}
=M*\delta S.
\label{A.8}
\end{equation}
Here, $\delta s_3(\infty) = \delta s_4(\infty) = s_5(\infty) = 0$. Taking the Fourier transforms of $\delta s_j(\tau)$ by $I_j(\omega) = \int_{0}^{\infty}d\tau e^{i(\omega-\omega_{\rm{p}})\tau}\delta s_j(\tau)$ with partial integration, we have
\begin{equation}
I(\omega) = [M + i(\omega-\omega_{\rm{p}})\mathbb{1}]^{-1}[\lim_{\tau\to\infty}\delta S(\tau)e^{i(\omega-\omega_{\rm{p}})\tau}-\delta S(0)].
\label{A.9}
\end{equation}
Because $\lim\limits_{\tau\to\infty}\delta S(\tau)e^{i(\omega-\omega_{\rm{p}})\tau}=0$,
\begin{equation}
I(\omega) = -[M+i(\omega-\omega_{\rm{p}})\mathbb{1}]^{-1}\delta S(0).
\label{A.10}
\end{equation}
Specifically, $I_3(\omega)$ is given by
\begin{eqnarray}
I_3(\omega)&=&\frac{\left | \overline s_1 \right |^2-\overline s_2}{\mu_1}\nonumber\\
&+&\frac{\Omega^2(\overline s_1^*)^2-\Omega^2(\left | \overline s_1 \right |^2-\overline s_2)\mu_2/\mu_1-2i\Omega\overline s_1^* \overline s_2\mu_2}{2\mu_1\mu_2\mu_3+\Omega^2(\mu_1+\mu_2)},\nonumber\\
\label{A.11}
\end{eqnarray}
where $\mu_1 = -\Gamma_2 + i\delta\omega_{01}$, $\mu_2 = -\Gamma_2 + i(\omega+\omega_{01} - 2\omega_{\rm{p}})$, and $\mu_3 = -\Gamma_1 + i\delta\omega_{01}$ with $\delta\omega_{01}=\omega-\omega_{01}$. Combining the above results, the incoherent part of the spectrum is obtained as
\begin{equation}
S_i(\omega) = \frac{\Gamma_r}{\pi}Re[I_3(\omega)],
\label{A.12}
\end{equation}
which is the same as Ref~\cite{koshino2012control}.

When the pump is on resonance with the qubit, if the pump power is strong ($\Omega\gg\Gamma_{1,2}$), $s_2\approx\frac{1}{2}$ and $s_1\approx\frac{-i\Gamma_1}{2\Omega}$. Then, $I_3(\omega)$ can be simplified to
\begin{eqnarray}
I_3(\omega)&\approx&-\frac{1}{4\mu_1}-\frac{1}{4\mu_1}\frac{\mu_3+\Gamma_1}{\mu_1\mu_3+\Omega^2}\nonumber\\
&\approx&-\frac{1}{4\mu_1}-\frac{1}{4}\left\{\frac{1}{\Gamma_s+i(\delta\omega_{01}+\Omega)}+\frac{1}{\Gamma_s+i(\delta\omega_{01}-\Omega)}\right\},\nonumber\\
\label{A.23}
\end{eqnarray}
where $\Gamma_{\rm{r}} = \Gamma_1 - \Gamma_{\rm{n}}$ and $\Gamma_{\rm{s}} = \frac{\Gamma_1 + \Gamma_2}{2}$. Therefore, Eq.~(\ref{A.12}) becomes
\begin{eqnarray}
S_i(\omega)&\approx&\frac{1}{\pi}\frac{\hbar\omega_{01}\Gamma_{\rm{r}}}{4}\left\{\frac{\Gamma_{\rm{s}}}{(\delta\omega_{01}+\Omega)^2+\Gamma_{\rm{s}}^2}\right.\nonumber\\
&+& \left.\frac{2\Gamma_2}{(\delta\omega_{01})^2+\Gamma_2^2}+\frac{\Gamma_{\rm{s}}}{(\delta\omega_{01}-\Omega)^2+\Gamma_{\rm{s}}^2}\right\}.
\label{A.13}
\end{eqnarray}

\section{Asymmetric Mollow triplet}\label{sec:Asymmetric}

In this section, we explain how dephasing leads to asymmetry in the off-resonant Mollow triplet. For the driven qubit, the states in the dressed-state basis can be written as
\begin{eqnarray}
\ket{n,+} &=& \sin \theta \ket{g,n+1} + \cos \theta \ket{e,n}, \\
\ket{n,-} &=& \cos \theta \ket{g, n+1} - \sin \theta \ket{e,n},
\end{eqnarray}
where $\ket{g}$ ($\ket{e}$) is the ground (excited) state of the qubit, $n$ is the number of drive photons, and $\theta$ is defined by
\begin{equation}
\tan 2 \theta = - \frac{\sqrt{\Delta^2+\Omega^2}}{\Delta}.
\end{equation}
A sketch of the dressed states is shown in Fig.~\ref{offresonatTheory}(a). To find the transition rates between the dressed states caused by relaxation, i.e., coupling of an environment to $\sigma_x$, we calculate the matrix elements
\begin{eqnarray}
\bra{n,+} \sigma_x \ket{n+1,+} &=& \sin \theta \cos \theta, \\
\bra{n,-} \sigma_x \ket{n+1,+} &=& \cos^2 \theta, \\
\bra{n,+} \sigma_x \ket{n+1, -} &=& -\sin^2 \theta, \\
\bra{n,-} \sigma_x \ket{n+1,-} &=& - \sin \theta \cos \theta.
\end{eqnarray}
Thus, Fermi's golden rule gives that the transition rates are
\begin{eqnarray}
 \Gamma_{++} &\propto& \sin^2 \theta \cos^2 \theta, \\
 \Gamma_{+-} &\propto& \cos^4  \theta, \\
\Gamma_{-+} &\propto& \sin^4 \theta, \\
\Gamma_{--} &\propto& \sin^2 \theta \cos^2 \theta.
\end{eqnarray}
In the case of resonant drive, $\Delta = 0$, we have $\theta = \pi /4$ and all the transition matrix elements are equal.

As illustrated in Fig.~\ref{offresonatTheory}(b), the transitions caused by relaxation are either between or within the $+$ and $-$ subspaces. Due to energy conservation, the product of the transition rate from the $+$ subspace to the $-$ subspace, $\Gamma_{+-}$, and the occupation probability of state this subspace, $P_{+} $, equals the product of the transition rate from the $-$ subspace to the $+$ subspace, $\Gamma_{-+}$, and the occupation probability of this subspace, $P_{-}$:
\begin{equation}
\Gamma_{+-} P_{+} = \Gamma_{-+} P_{-}.
\label{eq:DetailedBalance}
\end{equation}
If the drive is off-resonant, the transition rates are not the same. For $\delta < 0$, we have $\Gamma_{+-} > \Gamma_{-+}$, and for $\delta > 0$, we have $\Gamma_{-+} > \Gamma_{+-}$, i.e., the sideband that is closest to the qubit frequency has the highest transition rate. However, the emission spectrum is still symmetric, since the number of emitted photons in each sideband is given by the product the corresponding occupation probability and transition rate.

The presence of pure dephasing adds an additional term $H_\phi \propto \sigma_z (a + a^\dagger)$, where $a$ and $a^\dag$ are annihilation and creation operators for a bath, to the interaction Hamiltonian. The effect that this has on the dressed states can be understood by calculating the transition-matrix elements of $\sigma_z$ between the dressed states. We find
\begin{equation}
\bra{n,+} \sigma_z \ket{n,-} = -2 \sin \theta \cos \theta.
\end{equation}
All matrix elements of $\sigma_z$ for transitions between states with different number of drive photons are zero. The effect of pure dephasing is thus to cause transitions as sketched in Fig.~\ref{offresonatTheory}(a) and (b). Both upward and downward transitions are almost equally likely, since the corresponding transition energies are small compared to $k_B T$.

The pure dephasing thus modifies the condition for equilibrium from Eq.~(\ref{eq:DetailedBalance}) to
\begin{equation}
(\Gamma_{+-} + \Gamma_\phi) P_+ = (\Gamma_{-+} + \Gamma_\phi) P_-.
\end{equation}
This means that a non-zero $\Gamma_\phi$ pushes the state of the system closer to $P_{-} = P_{+}$ than it otherwise would have been. However, since the transition rates corresponding to relaxation remain the same as before, the result is that more photons are emitted at the frequency of the transition with the larger transition rate. This leads to an asymmetric power spectrum, where more photons are emitted in the sideband closest to the qubit frequency than in the sideband furthest away from the qubit frequency.

\section{Reflection coefficient}\label{sec:Reflection}
From the input-output relation, the output coherent field $\alpha_{\rm{out}}$ is the sum of the incoming coherent field $\alpha_{\rm{in}}$ and the field emitted by the atom:
\begin{equation}
\alpha_{\rm{out}} = \alpha_{\rm{in}} - i\sqrt{\Gamma_{\rm{r}}}\left \langle \sigma_-(t) \right \rangle,
\label{reflection}
\end{equation}
where $\alpha_{\rm{in}}=\frac{\Omega}{2\sqrt{\Gamma_r}}$. Combining this with Eq.~(\ref{A.6}), the reflection coefficient, $r = \frac{\alpha_{\rm{out}}}{\alpha_{\rm{in}}}$, becomes:
\begin{equation}
r = 1-\frac{i\Gamma_{\rm{r}}\Gamma_1(\Delta - i\Gamma_2)}{\Omega^2\Gamma_2 + \Gamma_1(\Delta^2 + \Gamma_2^2)}.
\label{B.1}
\end{equation}
In the case of a weak probe ($\Omega\ll \Gamma_2$), Eq.~(\ref{B.1}) becomes
\begin{equation}
r = 1-\frac{i\Gamma_{\rm{r}}}{\Delta + i\Gamma_2}.
\label{B.2}
\end{equation}
For a resonant probe ($\Delta=0$), Eq. (\ref{B.2}) is simplified to
\begin{equation}
r = 1-\frac{1}{\frac{\Omega^2}{\Gamma_1\Gamma_{\rm{r}}} + \frac{\Gamma_2}{\Gamma_{\rm{r}}}}.
\label{B.3}
\end{equation}

\section{Power Dissipation}\label{sec:Power}
At resonance ($\Delta = 0$), the input power is given by (setting $\hbar\omega_{01} = 1$)
\begin{equation}
P_{\rm{in}} = |\alpha_{\rm{in}}|^2 = \Omega^2/(4\Gamma_{\rm{r}}).
\label{C.1}
\end{equation}
The output power is a sum of coherent and incoherent contributions:
\begin{equation}
P_{\rm{out}} = P_{\rm{coh}} + P_{\rm{incoh}},
\label{C.2}
\end{equation}
with
\begin{equation}
P_{\rm{coh}} = P_{\rm{in}}|r|^2 = \frac{\Omega^2}{4\Gamma_{\rm{r}}}(1 - \frac{\Gamma_1\Gamma_{\rm{r}}}{\Omega^2 + \Gamma_2\Gamma_1})^2,
\label{C.3}
\end{equation}

\begin{eqnarray}
P_{\rm{incoh}}& = &\Gamma_{\rm{r}}(\langle\sigma_+\sigma_-\rangle-\langle\sigma_+\rangle\langle\sigma_-\rangle)\nonumber\\
& = &\frac{\Gamma_{\rm{r}}}{2}\frac{\Omega^2(\Gamma_1\Gamma_\phi + \Omega^2)}{(\Gamma_1\Gamma_2+\Omega^2)^2}.
\label{C.4}
\end{eqnarray}
In particular, when $\Gamma_\phi\ll\Gamma_1$, we have $P_{\rm{incoh}}\simeq 2\Gamma_r\rho_{11}^2$.
The net power loss is $P_{\rm{loss}} = P_{\rm{in}} - P_{\rm{coh}} - P_{\rm{incoh}}$.
\begin{equation}
P_{\rm{loss}} = \Gamma_{\rm{n}}\frac{\Omega^2}{2(\Gamma_1\Gamma_2 + \Omega^2)} = \Gamma_{\rm{n}}\rho_{11}.
\label{C.5}
\end{equation}
When $\Omega^2\gg\Gamma_1\Gamma_2$, the qubit is saturated. Then, we have $P_{\rm{in}}\approx P_{\rm{coh}}\approx\frac{\Omega^2}{4}$, $P_{\rm{incoh}}\approx\frac{\Gamma_{\rm{r}}}{2}$, and $P_{\rm{loss}}\approx\frac{\Gamma_{\rm{n}}}{2}$.


\bibliography{Characterization_Citation}


\end{document}